\documentclass[aps,prl,twocolumn,floatfix]{revtex4}
\usepackage{graphicx}
\usepackage{amsfonts}
\usepackage{amsmath,bm,amssymb}

\input{tcilatex}
\begin{document}

\title{Anisotropy, Itineracy, and Magnetic Frustration in High-T$_{C}$ Iron
Pnictides}
\author{Myung Joon Han}
\thanks{These authors have contributed equally to this paper.}
\author{Quan Yin$^{\ast }$}
\author{Warren E. Pickett}
\author{Sergey Y. Savrasov}
\affiliation{Department of Physics, University of California, Davis, California 95616, USA}
\date{\today }

\begin{abstract}
Using first-principle density functional theory calculations combined with
insight from a tight-binding representation, dynamical mean field theory,
and linear response theory, we have extensively investigated the electronic
structures and magnetic interactions of nine ferropnictides representing
three different structural classes. The calculated magnetic interactions are
found to be short-range, and the nearest ($J_{1a}$) and next-nearest ($J_{2}$%
) exchange constants follow the universal trend of $J_{1a}$/$2J_{2}\sim 1$,
despite their itinerant origin and extreme sensitivity to the z-position of
As. These results bear on the discussion of itineracy versus magnetic
frustration as the key factor in stabilizing the superconducting ground
state. The calculated spin wave dispersions show strong magnetic anisotropy
in the Fe plane, in contrast to cuprates.
\end{abstract}

\pacs{74.70.-b, 71.18.+y, 71.20.-b, 75.25.+z}
\maketitle

Recent discovery of the new high-temperature superconductor, LaO$_{1-x}$F$%
_{x}$FeAs with a transition temperature (T$_{C}$) of $26K$ \cite{JACS} has
triggered tremendous research activities on iron pnictides. Rare-earth (%
\textit{RE}) doping increases T$_{C}$ up to $55K$ for Sm \cite{Sm-43K,Sm-55K}%
. Replacing \textit{RE}-O layers with Li produces an intrinsic
superconductor LiFeAs with T$_{C}$ of $18K$ \cite{Tapp-Li}. The 122
ferropnictides, \textit{AL}Fe$_{2}$As$_{2}$ (\textit{AL}: Ca, Sr, Ba, K),
span another structural class with T$_{C}$ up to $38K$ \cite%
{Krellner-122,Jesche-122,Goldman-122,JZhao-Sr1,JZhao-Sr2,LZhao-122}. More
recently, arsenic-free FeSe$_{1-\delta }$ and Fe(Se$_{1-x}$,Te$_{x}$)$%
_{1-\delta }$ without any interlayer between Fe-(Se,Te) planes were found to
be superconducting at T$_{C}$ as high as $27K$ under pressure \cite%
{FeSe,FeSeTe-1,FeSeTe-2,FeSe-27K}. In spite of the accumulating reports of
both experiments and theories, the nature of the superconductivity and
magnetism is still far from clear. After several works have ruled out the
electron-phonon coupling \cite{Boeri,Haule}, and the coexistence of magnetic
fluctuation and superconductivity being confirmed by $\mu $SR \cite{Drew},
intensive investigations have been focused on the magnetic properties of
these systems \cite{Bao,Antropov,Ma,Si,Fang,Xu,Wu,Uhrig}. From the studies
up to now, one of the common and evident features is the interplay between
superconductivity and magnetism. It is clear, from the different structures,
that the essential physics lies in the iron plane forming the 2-dimensional
spin lattice.

Except for the Fe(Se,Te) family suggested to have different magnetic
structures by recent studies \cite{Bao,Antropov,Ma}, it is widely believed
that the first three classes of Fe pnictides have a common superconducting
mechanism closely related to magnetic interactions. In order to clarify the
raised issues and lead to further understanding, it is of key importance to
investigate the exchange interactions across different classes of compounds
and examine any trend or common features. However, material-specific
information of magnetic interactions is scarce in spite of active research
efforts. The direct probe of spin dynamics is inelastic neutron scattering,
which has been recently performed for SrFe$_{2}$As$_{2}$ \cite{JZhao-Sr2}
and CaFe$_{2}$As$_{2}$ \cite{McQueeney}. They have revealed that the
combination of nearest and next nearest neighbor exchange interactions $%
|J_{1a}+2J_{2}|$ is about $100meV$, but detailed data from individual
contributions, as well as their anisotropy and the proximity of the ratio $%
J_{1a}/2J_{2}$ to unity, which has been discussed extensively in recent
publications \cite{Si,Fang,Xu}, are still missing.

In this Letter, using first-principle linear response calculations \cite%
{RMP-1, RMP-2}, we provide the data of in-plane magnetic exchange couplings
for several Fe-based superconductors, and discuss their spin wave
dispersions. The data bear on the question of whether the values of exchange
constants indicates magnetic fluctuations play an important role. A total of
nine materials have been studied: \textit{RE} FeAsO (\textit{RE}: La, Ce,
Pr, Nd), \textit{AL}Fe$_{2}$As$_{2}$ (\textit{AL} : Ca, Sr, Ba, K), and
LiFeAs. Exchange interactions of these systems are found to be short-range
despite the metallic density-of-states (DOS), and the calculated interaction
strengths follow the universal behavior of $J_{1a}\approx 2J_{2}$ for all
materials, a relation that arises independently in the frustrated magnetic
picture \cite{Si,Fang,Xu}. Considering not only the variety of the materials
studied here but also the high sensitivity of the Fe moment to the
z-position of As atom \cite{ZPYin,Yildirim}, this universal behavior of the
exchange interactions is impressive. The calculated spin-wave dispersion
shows an anisotropic spin interaction which is different from the cuprates.

There have been several published tight-binding (TB) parametrizations of the
electronic structure of prototypical LaOFeAs in the vicinity of the Fermi
level using fits based either on Wannier functions or atomic basis sets \cite%
{TB-1,TB-2,TB-3,TB-4}. However the current situation still looks complicated
because the projected DOS deduced from electronic structure calculations are
based on the spherical harmonic projectors within the atomic spheres that
may not be very well suited for the extended Fe and As orbitals presented
here. Due to these complications even the crystal field splitting of Fe $d$
level appears to be controversial in the current literature \cite%
{TB-1,TB-2,TB-3,TB-4}. 
\begin{figure}[tbp]
\centering\includegraphics[width=9cm]{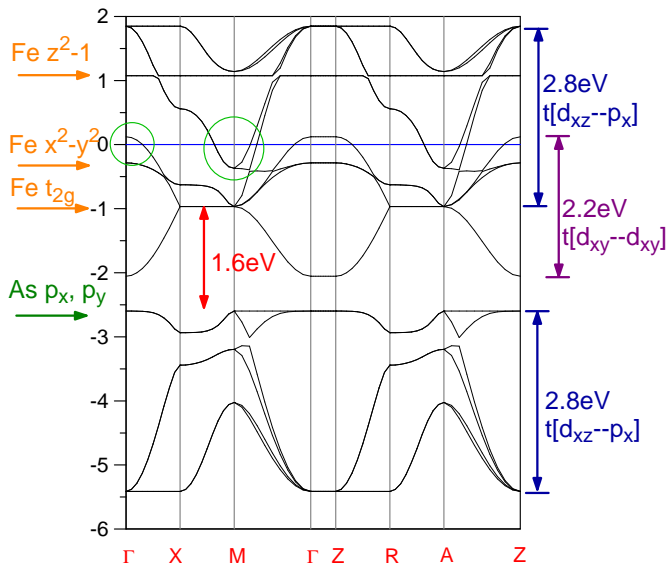}
\caption{(Color online) The tight-binding band structure of LaFeAsO. The
circles at the Fermi level on the $\Gamma $ and $M$ points indicate the hole
and electron pockets, respectively.}
\label{fig-TB}
\end{figure}

To better understand the complicated electronic structure around Fermi
level, we performed TB analysis by considering $d_{xz}$ and $d_{yz}$
orbitals of Fe $t_{2g}$ manifold hybridizing with the arsenic $p_{x}$ and $%
p_{y}$, respectively. As shown in Fig.~\ref{fig-TB}, the separation between
the energy levels of Fe-$t_{2g}$ and As-$p_{x,y}$ states is about $1.6eV$.
Accounting for the hybridization matrix element between $d_{xz}$-$p_{x}$ , $%
d_{yz}$-$p_{y}$ states, which is of the order of $1.8eV$, produces bonding
and antibonding bands, both having the bandwidth of $2.8eV$ with the Fermi
level falling into the antibonding part of the spectrum (approximately $1eV$
above the Fe $t_{2g}$ level). We also take into account the $d_{xy}$ state
of Fe which hybridizes with itself (hopping integral is approximately $0.3eV$%
), which produces an additional bandwidth of $2.2eV$. The resulting
bandwidth of Fe $d$-electron character near the Fermi level becomes $%
2.8+2.2/2=3.9eV$ as exactly seen in the LDA calculation \cite{Singh}. The
coordinate system used for this TB description is the original
crystallographic lattice where the spin alternates in the $(\pi ,\pi )$
direction. In this picture, the $\Gamma $-centered hole pockets (small
circle in Fig.~\ref{fig-TB}) are mostly of $d_{xy}$ character, and the $M$%
-centered pockets (large circle) are of $d_{xz}$, $d_{yz}$ character. This
picture can be fine-tuned further by including the $d_{x^{2}-y^{2}}$ state
which lies $0.3eV$ below the Fermi level and hybridizes primarily with As-$%
p_{x,y}$ states (hopping integral is about $0.8eV$) as well as hybridization
between $d_{xz,yz}$ orbitals with As $p_{z}$ states (hopping integral is
about $0.4eV$). Note that in this picture the Fe $d_{z^{2}-1}$ orbital
becomes unoccupied and lies $1eV$ above the Fermi level.

\begin{figure}[tbp]
\centering \includegraphics[width=5cm]{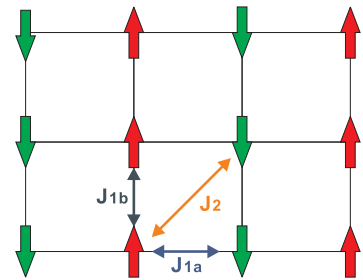}
\caption{(Color online) Spin arrangment and exchange interactions in the Fe
plane of the striped Q$_{m}$-AFM phase. The arrows on lattice sites indicate
the Fe spin directions.}
\label{QM-str}
\end{figure}

Now we discuss the exchange interactions. To calculate the interactions
between Fe moments, we used linear response theory \cite{J-RKKY,Wan} based
on first-principle density functional theory (DFT) calculations, which has
been successfully applied to the $3d$ transition-metal oxides and the $5f$
actinides metallic alloys \cite{Wan,Han}. We used the full potential
linearized muffin-tin orbital (LMTO) as the basis set \cite{Sav-96} and
local spin density approximation (LSDA) for the exchange-correlation (XC)
energy functional. The LSDA is fairly good to describe the itinerant Fe $3d$
states in these materials as shown in the previous studies and comparisons
with angle resolved photoemission\textbf{\ }\cite%
{ZPYin,Yildirim,Singh,LOFP-Nature}. In the calculations of \textit{RE}OFeAs
compounds, we used\ the LSDA+DMFT method \cite{RMP-1,RMP-2} in which the 
\textit{RE} $4f$ orbitals are treated as the localized ones within Hubbard I
approximation. $U=6eV$ and $J_{H}=0.86eV$ were used as the on-site Coulomb
repulsion and Hund's rule exchange parameter. Lattice constants are taken
from experiments, and we performed the calculations at various $z(As)$,
including experimental $z(As)_{\exp }$ and LDA optimized $z(As)_{LDA}$.

Fig.~\ref{QM-str} shows the spin structure of the Fe plane which is common
to the all these materials. From here on we use the $(\pi ,0)$ striped AFM
coordinate system, which is convenient to discuss the spin wave dispersions.
Magnetic interactions between Fe moments are governed by two dominating AFM
couplings $J_{1a}$ and $J_{2}$, and the FM nearest-neighbor exchange $J_{1b}$
is small. We found the exchange couplings $J(\mathbf{q})$ can be expressed
in terms of short-range exchange constants. This character suggests pursuing
a comparison with local moment models with AFM spin interactions\cite%
{Si,Fang,Xu}. The short range couplings do not conflict with the itinerant
magnet picture because although the Fe $3d$ orbital has finite DOS at the
Fermi level, the magnetic interactions can still remain short range, which
possibly reflects the bad metallicity and some correlation effects.

The calculated Fe magnetic moments and exchange interactions are summarized
in Table~\ref{tab}. We use the convention that positive $J$ means AFM
couplings. The calculated moments are consistent throughout the materials.
The calculations done at experimental $z(As)_{\exp }$ are known to predict
the moments about twice as large as experimental values, while at optimized $%
z(As)_{LDA}$ they give smaller moments. The cases in which DFT overestimates
magnetic moments are rare, and the cause is still under debate for Fe
oxypnictides. Although some theorists suggest it is due to the frustrated
magnetic structure \cite{Si}, Mazin and Johannes suggest an alternative
picture \cite{Mazin} based on magnetic fluctuation and inhomogenieties.
Importantly, the electronic structure features such as electron-hole
symmetry and the exchange interaction strengths are better described with $%
z(As)_{exp}$ when compared to available experimental data \cite%
{McQueeney,JZhao-Sr2}. Thus our discussion will be based on the results from 
$z(As)_{exp}$. The sensitivity of moments and exchange interactions to $%
z(As) $ is large. For example, in LaFeAsO the change of $z(As)$ by $0.04%
\mathring{A}$~ ($\Delta z(As)=0.005$ in terms of internal coordinates)
induces about $10\%$ difference in the moment and up to $20\%$ in the
exchange interactions \cite{ZPYin}. The same order of sensitivity was also
reported for CaFe$_{2}$As$_{2}$ \cite{Yildirim}. Therefore the deviation of
up to $8\%$ for moments and $30\%$ for major exchange interactions ($J_{1a}$
and $J_{2}$) are not significant, and become much smaller if $z(As)$ could
be refined for each material. Taking this into account, we can say that the
magnetic moments and exchange interactions are uniform throughout the
materials considered here.

\begin{table}[tbp] \centering%
\begin{tabular}{lcccccc}
\hline\hline
System & Moment & $J_{1a}$ & $J_{2}$ & $J_{1b}$ & $J_{1a}/2J_{2}$ & $%
J_{1a}+2J_{2}$ \\ \hline
LaFeAsO & $1.69$ & $47.4$ & $22.4$ & $-6.9$ & $1.06$ & $92.2$ \\ 
CeFeAsO & $1.79$ & $31.6$ & $15.4$ & $2.0$ & $1.03$ & $62.4$ \\ 
PrFeAsO & $1.76$ & $57.2$ & $18.2$ & $3.4$ & $1.57$ & $93.6$ \\ 
NdFeAsO & $1.49$ & $42.1$ & $15.2$ & $-1.7$ & $1.38$ & $72.5$ \\ \hline
CaFe$_{2}$As$_{2}$ & $1.51$ & $36.6$ & $19.4$ & $-2.8$ & $0.95$ & $75.4$ \\ 
SrFe$_{2}$As$_{2}$ & $1.69$ & $42.0$ & $16.0$ & $2.6$ & $1.31$ & $74.0$ \\ 
BaFe$_{2}$As$_{2}$ & $1.68$ & $43.0$ & $14.3$ & $-3.1$ & $1.51$ & $71.5$ \\ 
KFe$_{2}$As$_{2}$ & $1.58$ & $42.5$ & $15.0$ & $-2.9$ & $1.42$ & $72.5$ \\ 
\hline
LiFeAs & $1.69$ & $43.4$ & $22.9$ & $-2.5$ & $0.95$ & $89.2$ \\ \hline\hline
\end{tabular}
\caption{Calculated Fe moments (in $\mu _{B}$) and in-plane exchange interactions (in 
$meV$), using experimental $z(As)$.}\label{tab} 
\end{table}%

One of the most important quantities to understand the magnetism and
possibly the superconducting mechanism in these materials is the ratio of $%
J_{1a}/2J_{2}$ , which has so far not been measured nor calculated.
According to the spin Hamiltonian models \cite{Si,Fang,Xu}, assuming Fe
pnictides as magnetic Mott insulators like cuprates, at $J_{1a}/2J_{2}%
\approx 1$ the system is close to the quantum critical regime, so a
superconducting ground state may appear as a result of the magnetic
fluctuation \cite{Si,Fang,Xu,Chubukov}. Note that the calculated ratios
shown in Table~\ref{tab} are all around unity, demonstrating that this
universal behavior of $J_{1a}/2J_{2}$ can arise from itinerant magnetism,
without the system being close to a Mott transition. The deviations of $%
J_{1a}/2J_{2}$ from unity reflect not only the intrinsic material properties
but also the sensitive dependence on $z(As)$. Although there is no apparent
relation between the $J_{1a}/2J_{2}$ ratio and T$_{C}$, the universal
feature of $J_{1a}/2J_{2}$ near unity is closely associated to
superconductivity since it is present throughout the materials studied here.
The connection between itinerant AFM and superconductivity has been
discussed previously \cite{Moriya}.

Another important quantity is $\left\vert J_{1a}+2J_{2}\right\vert $ which
determines the spin wave velocity in the $(\pi ,0)$ direction, and can be
directly probed by neutron scattering experiments. The available
experimental data are in general agreement with our calculation. For SrFe$%
_{2}$As$_{2}$ calculation shows $\left\vert J_{1a}+2J_{2}\right\vert =74meV$%
, not much smaller than the $100\pm 20meV$ measured by neutron scattering 
\cite{JZhao-Sr2}. Also, for CaFe$_{2}$As$_{2}$ our calculated $\left\vert
J_{1a}+2J_{2}\right\vert =75meV$ is slightly smaller than the measured $%
95\pm 16meV$ \cite{McQueeney} (derived from the observed spin wave velocity,
see eq.(\ref{swvp}) below). Especially our result for BaFe$_{2}$As$_{2}$ is
in good agreement with recent experiment by Ewings \textit{et al.} \cite%
{Ewings}: One of their best fits shows that $J_{1a} = 36 meV, J_2 = 18 meV$,
and $J_{1b} = -7 meV$.

\begin{figure}[tbp]
\centering \includegraphics[width=8cm]{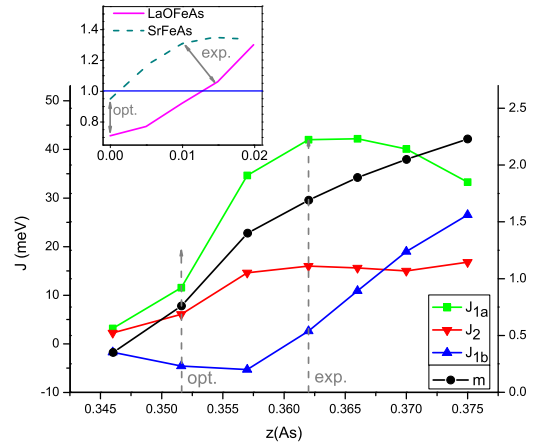}
\caption{value (exp.) and theoretically optimized value (opt.) of $z(As)$.
In the inset figure, the LDA-optimized z(As) is set to be zero reference.}
\label{fig-J_z}
\end{figure}

As an example, Fig.~\ref{fig-J_z} shows the $z(As)$-dependence of the
magnetic moments and interactions of SrFe$_{2}$As$_{2}$. The moment is a
simple monatomic function of $z(As)$ ranging from $0.35\mu _{B}$ to $2.23\mu
_{B}$. The three $J$'s have different behaviors. $J_{1a}$ increases rapidly
with $z(As)$ at the beginning, saturates in the middle, and eventually turns
down. This behavior is a result of the hybridization between Fe $d_{xz,yz}$
and As $p_{x,y}$ orbitals, as we discussed in the tight-biding
representation. Due to the shape and orientation of the Fe $d_{xz,yz}$ and
As $p_{x,y}$ orbitals, there is a certain $z(As)$ that gives the maximum
overlapping, and hence largest $J_{1a}$. Note that $J_{1b}$ changes sign at $%
z(As)_{\exp }$, and eventually surpasses $J_{2}$. Also, $J_{1a}$ and $J_{2}$
plateau in the small region around $z(As)_{exp}$. Similar behaviors are also
found in other materials. The $J_{1a}/2J_{2}$ and $\left\vert
J_{1a}+2J_{2}\right\vert $ values presented in Table~\ref{tab} are robust
against the small deviations in $z(As)$ around the experimental values. From
the data one can also calculate $J_{1a}/2J_{2}$ vs. $z(As)$, which reveals
the existence of the \textquotedblleft sweet spot" where the optimal ratio $%
J_{1a}/2J_{2}=1$ is achieved independent of any Heisenberg model assumption.
In the case of SrFe$_{2}$As$_{2}$ it is $z(As)=0.357$ (the inset of Fig.~\ref%
{fig-J_z}).

\begin{figure}[tbp]
\centering \includegraphics[width=9cm]{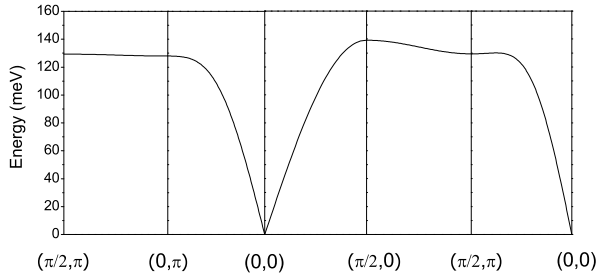}
\caption{The calculated spin wave dispersion of SrFe$_{2}$As$_{2}$ along
high-symmetry lines, the exchagne constants are given in Table~\protect\ref%
{tab}.}
\label{fig-SW}
\end{figure}

The calculated spin wave dispersion gives more intuitive information about
the magnetic interaction and anisotropy of these systems \cite%
{Uhrig,JZhao-Sr2}. The dispersion relation of the 2D striped-AFM lattice
reads 
\begin{equation}
\omega (\mathbf{q})=S\sqrt{(\mathcal{J}_{0}+J_{1b}(\mathbf{q}))^{2}-(J_{1a}(%
\mathbf{q})+J_{2}(\mathbf{q}))^{2}}.  \label{SWD}
\end{equation}%
Using the calculated magnetic exchange constants, we plot the spin wave
dispersion of SrFe$_{2}$As$_{2}$ in Fig. \ref{fig-SW}, whose $S=0.94$ is
taken from experiment \cite{JZhao-Sr1}. The non-symmetric dispersions in $%
(0,0)-(0,\pi )$ and $(0,0)-(\pi ,0)$ directions indicate in-plane magnetic
anisotropy, which is a major difference from cuprates. At small $q$ near $%
(0,0)$, the spin wave velocity in the $(\pi ,0)$ direction is 
\begin{equation}
v_{\perp }=2aS\left\vert J_{1a}+2J_{2}\right\vert ,  \label{swvp}
\end{equation}%
which is the relation used to experimentally determine $\left\vert
J_{1a}+2J_{2}\right\vert $, such as for SrFe$_{2}$As$_{2}$ \cite{JZhao-Sr2}.
The difference in $J_{1a}$ and $J_{1b}$, a direct consequence of the Q$_{M}$%
-AFM ordering that breaks in-plane symmetry, accounts for the anisotropy in $%
(\pi ,0)$ and $(0,\pi )$ directions. These anisotropic spin waves can be
directly probed by neutron scattering experiments on single crystals.

To conclude, we have studied magnetic exchange interactions in various
Fe-based high T$_{C}$ superconductors using first-principle based linear
response calculations. From the nine different materials, the magnetic
interactions are short-range and can be well described by the first and
second nearest-neighboring interactions. Importantly $J_{1}/2J_{2}$ is close
to unity for all the cases, just as would be the case for the frustration
limit of a local moment model. Calculated spin wave dispersions show the
magnetic anisotropy and the roles of the three in-plane exchange
interactions. Our result strongly suggests the magnetic fluctuation as the
pairing mechanism for the superconducting ground state.

The authors would like to thank Elihu Abrahams (Rutgers Univ.) and Rajiv
Singh (UC Davis) for helpful discussions on the spin wave dispersion. The
authors acknowledge support from NSF Grants DMR-0606498 (S.Y.S) and
DMR-0421810 (W.E.P), and collaborative work supported by DOE SciDAC Grants
SE-FC0206ER25793 and DE-FC02-06ER25794.

\end{document}